\documentclass[prb,twocolumn,showpacs,amsmath,amssymb,superscriptaddress]{revtex4-1}
\usepackage{graphicx}
\usepackage{dcolumn}
\usepackage{bm}
\usepackage{amsmath}
\usepackage{amsmath}

\newcommand{\nc}{\newcommand}

\nc{\be}{\begin{equation}} \nc{\ee}{\end{equation}}
\nc{\bea}{\begin{eqnarray}} \nc{\eea}{\end{eqnarray}}
\nc{\bean}{\begin{eqnarray*}} \nc{\eean}{\end{eqnarray*}}

\begin{document}

\title{Spin dynamics in the strong spin-orbit coupling regime}
\author{Xin Liu}
\affiliation{ Department of Physics, Texas A\&M University, College
Station, TX 77843-4242, USA}
\author{Xiong-Jun Liu}
\affiliation{ Department of Physics, Texas A\&M University, College
Station, TX 77843-4242, USA}
\author{Jairo Sinova}
\affiliation{ Department of Physics, Texas A\&M University, College
Station, TX 77843-4242, USA}
\affiliation{Institute of Physics ASCR, Cukrovarnick\'a 10, 162 53 Praha 6, Czech
Republic }
\date{\today}

\begin{abstract}
We study the spin dynamics in a high-mobility two dimensional electron gas (2DEG) with generic spin-orbit interactions (SOIs). 
We derive a set of spin dynamic equations which capture the  purely exponential to the damped oscillatory spin evolution modes 
observed in different regimes of SOI strength. 
Hence we provide a full treatment of the D'yakonov-Perel's mechanism by using the microscopic linear response theory from the 
weak to the strong  SOI limit. We show that the damped oscillatory modes appear when the electron scattering time is larger than half of the spin precession time due to the SOI, in agreement with recent observations. We propose a new way to measure the scattering time and the relative strength of Rashba and linear Dresselhaus SOIs based on these modes and optical grating experiments. We discuss the physical interpretation of each of these modes in the context of Rabi oscillation. 
\end{abstract}

\maketitle

\section{Introduction}\label{1}
In recent years research in semiconductor based devices has  incorporated the spin degree of freedom 
as a new state variable in novel electronic devices with potential for future applications. The SOI is a key tool to electrically manipulate the spin and realize such devices. However, the SOI is a double-edged sword because it will also induce random spin precession through an angle $\Omega_{so}\tau$ between collisions with impurities, where $\tau$ is the electron life time. This is known as the  D'yakonov-Perel's(DP)  mechanism\cite{DP-1,DP-2,DP-3} and  dominates the spin relaxation in the technologically important III-V semiconductors.\cite{Meier} Therefore it is very important to understand fully the DP mechanism for the possible application and further development of spintronics devises. Although the study of DP mechanism in semiconductors in the presence of SOI was initiated long ago, most of the theoretical research \cite{burkov,Halperin,a+b,e-e,Wu,DHLee} focuses on the weak spin-orbit coupling (SOC) regime  where $\Omega_{so}\tau \ll 1$. However, as high-mobility 2DEG systems are created, it is now not difficult to reach the strong SOI regime experimentally where $\Omega_{so}\tau >1$  at low temperatures as long as the mobility is approximately larger than $1.2 \times 10^{5} {\rm cm^2 /Vs}.$\cite{exq=0-1} The spin evolution in this regime is observed to be damped oscillations in the uniform \cite{exq=0,exq=0-1} and nonuniform spin polarized system,\cite{nature2005,prl2007,psh} which can not be described by spin-charge drift-diffusion equations derived for the weak SOC regime and lacks a clear theoretical explanation.

Here, we study the spin dynamics theoretically from the weak to strong SOC regime. The method we use is linear response theory.\cite{diffth,a+b,burkov} We derive a set of spin dynamical equations in the uniform spin polarized 2DEG with different SOIs. We show analytically that for $\Omega_{so}\tau >\frac{1}{2}$ the damped oscillations appear. The decay rate in this case is proportional to $\frac{1}{\tau}$ instead of $\tau$ as in the weak SOC regime. The cubic Dresselhaus term is shown to reduce the oscillatory frequency and increase the decay rate in the strong SOC regime. The spin dynamics for non-uniform spin polarization with spatial frequency $q$ in the strong SOC regime is obtained by solving the equations numerically. We discuss these dynamics by using the analogy with  Rabi oscillations between two momentum states which are gaped by the SOI. Our results match the experimental observation quantitatively. We also show how to exploit our analysis to create an accurate measurement of the strength of Rashba and linear Dresselhaus SOIs in a 2DEG, hence allowing a full characterization of different device samples which will lead to a more accurate modeling and predictability of the optimal operating physical regimes. 

\section{Model Hamiltonian and density matrix response function}

Normally in the 2-D semiconductor heterostructures, we have three kinds of SOIs, namely the linear Rashba \cite{Rashba-1,Rashba-2} term and the linear and cubic Dresselhaus \cite{Dresselhaus} terms.
The Hamiltonian takes the form
\begin{eqnarray}\label{ha}
H=\frac{k^2}{2m}+\mathbf{h(k)} \cdot \hat{\sigma},
\end{eqnarray}
where $\mathbf{h(k)}$ is the effective magnetic and contains Rashba, linear and cubic Dresselhaus terms which are
\begin{eqnarray}\label{soi}
\mathbf{h}^R(\mathbf{k})=\alpha (-k_y,k_x),\\
\mathbf{h}^{D_1} (\mathbf{k})=\beta_1 (k_y,k_x),\\
\mathbf{h}^{D_3}(\mathbf{k})=-2\beta_3\cos2\theta(-k_y,k_x),
\end{eqnarray}
where $k_f$ is the Fermi wave vector. Here we take $\theta$ as the angle between the wave vector $\mathbf{k}$ and the $[110]$ direction which is the $x$ axis in our coordinates. The above SOIs split the spin-degenerate bands and dominate the spin dynamics in the 2DEG. The corresponding SOC Hamiltonian and the spin precession frequency $\Omega_{so}$ takes the form:
\begin{eqnarray}\label{hso}
H^{so}&=&(\lambda_1-2\beta_3\cos2\theta)k_x\sigma_y+(\lambda_2+2\beta_3\cos2\theta)k_y\sigma_x, \nonumber \\
\end{eqnarray}
where
$\lambda_1=\alpha+\beta_1$, $\lambda_2=\beta_1-\alpha$.

We derive the spin dynamic equations from the density matrix response function,\cite{diffth} used previously in Refs. \onlinecite{burkov,a+b} in the weak spin-orbit coupling regime $\Omega_{so}\tau \ll 1$. The spin diffusion is dominated by the pole of the spin-charge diffusion propagator or "diffuson":
\begin{eqnarray}\label{diffuson-1}
\mathcal{D}=[1-\hat{I}]^{-1}
\end{eqnarray}
and
\begin{eqnarray}\label{II}
\hat{I}_{\sigma_1\sigma_2,\sigma_3\sigma_4}=\frac{1}{2m\tau}\int \frac{d^2k}{(2\pi)^2}G^A_{\sigma_3\sigma_1}(k,0)G^R_{\sigma_2\sigma_4}(k+q,\Omega),\nonumber \\
\end{eqnarray}
where $\sigma_i$ is just a number which can be $1$ or $2$.\cite{burkov} Here, we will calculate the diffuson matrix exactly and make it valid in any SOC regime. It is more convenient to write the Eq. \ref{II} in a classical charge-spin space 
\begin{eqnarray}\label{Ics}
I^{\alpha\beta}={\rm Tr}(\sigma_{\alpha} \hat{I} \sigma_{\beta}),
\end{eqnarray}
where $\alpha,\beta=c,x,y,z.$\cite{burkov}
\section{Uniform spin polarization}
In the case of a uniform spin polarized 2DEG system, i.e. $q=0$, because the effective magnetic field due to the SOI has inversion symmetry in momentum space, only the diagonal elements of the diffuson matrix are nonzero, which means the spin $x$, $y$, $z$ and charge are not coupled to each other. Therefore, when considering the uniform spin polarization along the $z$ direction, only $I^{zz}$ needs to be calculated. First, we neglect the cubic Dresselhaus term which is normally much smaller than the linear Dresselhaus term. We find the pole of diffusion matrix by solving the equation
\begin{eqnarray}\label{Izz}
1-I^{zz}=1-\frac{1-i\Omega\tau}{\sqrt{((1-i\Omega\tau)^2+(\Omega_{so}\tau)^2)^2-\gamma^2 (\Omega_{so} \tau)^4}}=0,\nonumber\\
\end{eqnarray}
where $\Omega$ is the frequency of the spin evolution, $\Omega_{so}=2\sqrt{\alpha^2+\beta_1^2}k_f$, $\gamma=\frac{2\alpha\beta_1}{\alpha^2+\beta_1^2}=\frac{\lambda_1^2-\lambda_2^2}{\lambda_1^2+\lambda_2^2}$,
$k_f$ is the Fermi wave vector and $I^{zz}$ is obtained from the exact angular integration of Eq.(\ref{II}). The details of calculating $I^{zz}$ are shown in the appendix. The eigenmode of the spin dynamical evolution takes the form
\begin{widetext}
\begin{eqnarray}\label{zmode}
i\Omega \tau =\frac{1}{2} \left(2 - \sqrt{2} \sqrt{1-2 (\Omega_{so}\tau)^2 +\sqrt{1-4 (\Omega_{so}\tau)^2+4 (\Omega_{so}\tau)^4 \gamma^2}}\right).
\end{eqnarray}
\end{widetext}
Note that $\gamma\leq 1$ and 
\begin{eqnarray}
(1-4 (\Omega_{so}\tau)^2+4 (\Omega_{so}\tau)^4 \gamma^2)\leq (1-2\Omega_{so}^2\tau^2). \nonumber
\end{eqnarray}Therefore, as long as $1-4 (\Omega_{so}\tau)^2+4 (\Omega_{so}\tau)^4 \gamma^2<0$, a non-equilibrium spin polarization will exhibit damped oscillation with respect to time.
\begin{figure}[htbp]
\includegraphics[width=1.0\columnwidth]{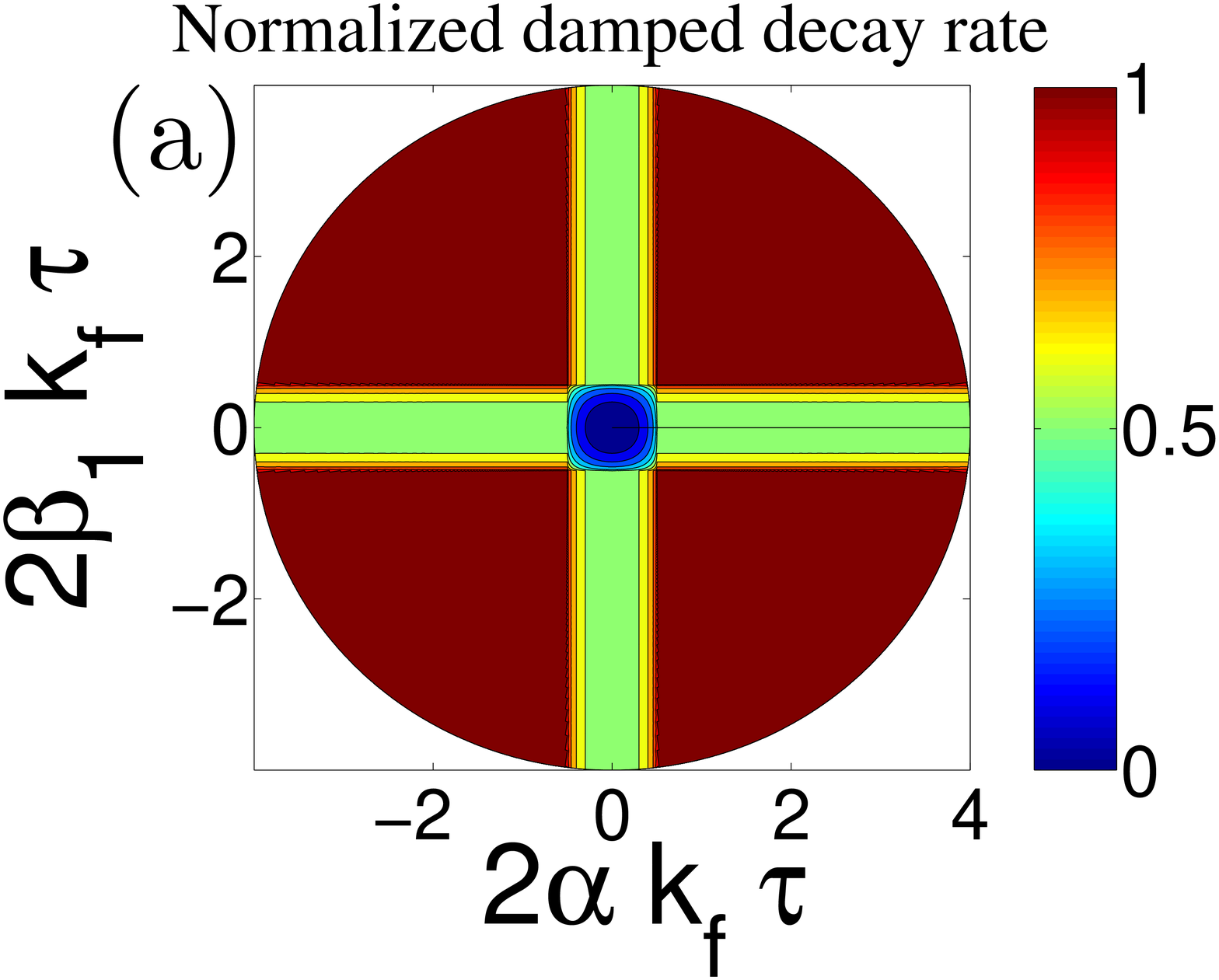} \\
\includegraphics[width=1.0\columnwidth]{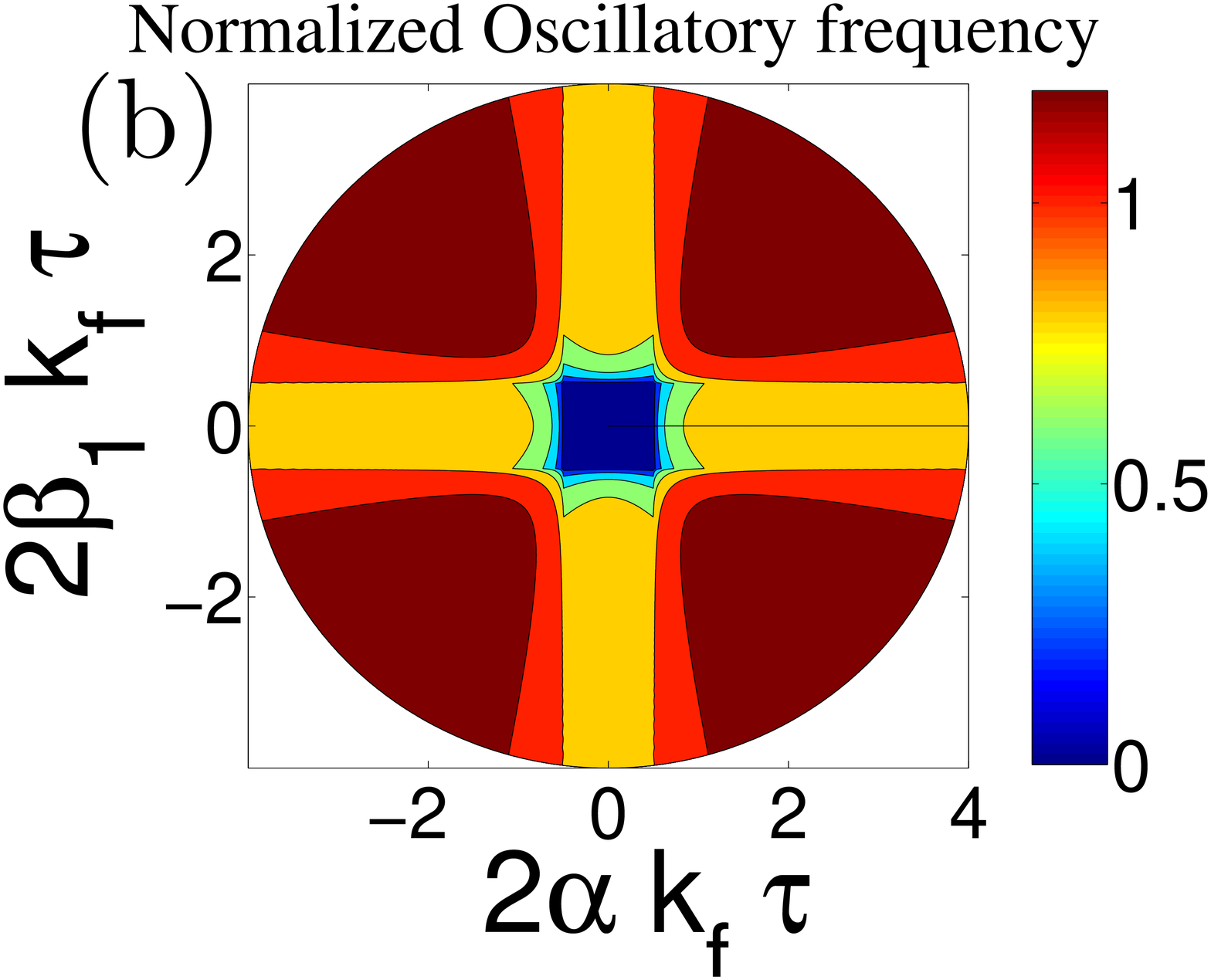}
\caption{The uniform spin dynamics from the weak to the strong spin-orbit coupling regime in the presence of both Rashba and linear Dresselhaus terms. (a) The normalized exponential decay rate $\Im{(\Omega \tau)}$ is shown as a function of normalized Rashba and linear Dresselhaus SOI. (b) The nonzero normalized oscillatory frequency, $\frac{\Re{(\Omega)}}{\Omega_{so}}$, is nonzero whenever  $2\alpha k_f \tau \geq \frac{1}{2}$ or $2\beta_1 k_f \tau \geq \frac{1}{2}$. }
\label{disp}
\end{figure}

In the case of $\alpha=0$ or $\beta_1=0$, the eigenmode takes the form
\begin{eqnarray}\label{puresoi}
i\Omega\tau=\frac{1}{2}-\frac{1}{2}\sqrt{1-4\Omega^2_{so}\tau^2}.
\end{eqnarray}
When $\Omega_{so}\tau>1/2$, the decay rate changes from the exponential decay mode to the damped oscillation mode. The oscillatory frequency in the clean limit, $\tau\rightarrow \infty$, is $\Omega_{so}$. Several experiments\cite{exq=0,exq=0-1,nature2005,prl2007} observe the damped oscillation mode of spin evolution at low temperature. However their analysis did not explain quantitatively when this kind of mode appears but just qualitatively argue that it appears in the regime where $\Omega_{so}\tau>1$. Our theory agrees with the recent experiment \cite{exq=0} in which the authors observe that when the temperature is above $5$ K, the oscillation will disappear. In their system this corresponds to $\Omega_{so}\tau_p^*\approx0.48$, which is close to our result $1/2$. Here $\tau_p^*$ is different to the transport scattering time $\tau_p$ obtained from the mobility; this difference is due to the Coulomb interaction effect on spin-currents and spin dephasing.\cite{nature2005,e-e} This e-e interactions treatment is beyond our paper and will not be discussed in this work. The $\tau$ here is corresponding to $\tau_p^*$. When the oscillatory mode appears, the damped decay rate is always equal to $\frac{1}{2\tau}$  when either $\alpha=0$ or $\beta_1=0$. This result matches  the recent experiment \cite{exq=0-1} in which the authors found the decay rates for several different 2DEGs  always equals  $\frac{1}{1.9\tau}$ when the damped oscillatory mode appears, in agreement with our theoretical result.

As the linear and cubic Dresselhaus terms always coexist, we have to consider the effect of cubic Dresselhaus term on Eq (\ref{puresoi}). We do this in the simplest case, when Rashba coefficient is zero. In this case, the diffuson matrix element $I^{zz}$ takes the form
\begin{eqnarray}\label{izz-cubic-5}
I^{zz}&=&\frac{1-i\Omega\tau}{\sqrt{(1-i\Omega\tau)^2+\Omega_{so}^2\tau^2(1+2(\frac{\beta_3}{\beta_1})^2-2\frac{\beta_3}{\beta_1})}}\nonumber \\
&&\times \frac{1}{\sqrt{(1-i\Omega\tau)^2+\Omega_{so}^2\tau^2}}
\end{eqnarray}
where $\Omega_{so}=2\beta_1 k_f$ and $\delta=2\frac{\beta_3}{\beta_1}(1-\frac{\beta_3}{\beta_1})$.
The corresponding spin decay rate is
\begin{eqnarray}\label{izz-cubic-6}
&&i\Omega\tau=1-\nonumber \\
&&\frac{\sqrt{(1+\sqrt{1-4\Omega_{so}^2\tau^2+2\Omega_{so}^2\tau^2\delta+\Omega^4_{so}\tau^4\delta^2})^2-\Omega^4_{so}\tau^4\delta^2}}{2}.\nonumber \\
\end{eqnarray}
Eq (\ref{puresoi},\ref{izz-cubic-6}) show that the cubic term will increase the exponential decay rate and decrease the oscillatory frequency. To show the effect of the cubic Dresselhaus term,  the real(imagine) value of the damped oscillatory frequency when $\beta_3\neq 0$ is divided by the value when $\beta_3=0$. This ratio is ploted in Fig \ref{disp} with respect $\beta_3/\beta_1$ and $2\beta_1\tau$. When $\frac{\beta_3}{\beta_1}<0.2$,  the effect of  the cubic term is very small and can be neglected.  In this case, the damped decay rate is always equal to $\frac{1}{2\tau}$ as long as $\Omega_{so}>\frac{1}{2}$ and the oscillatory frequency $\Omega$ approach $\Omega_{so}$ when $\Omega_{so}\tau \gg 1$. This provides a reliable way to measure the momentum scattering time $\tau$.  Further, the strength of the linear Dresselhaus SOI can be obtained from Eq. \ref{puresoi} once we know $\tau$ and the oscillatory frequency from the measurements. These will be discussed in a later section.
\begin{figure}[ht]
\includegraphics[width=1.0\columnwidth]{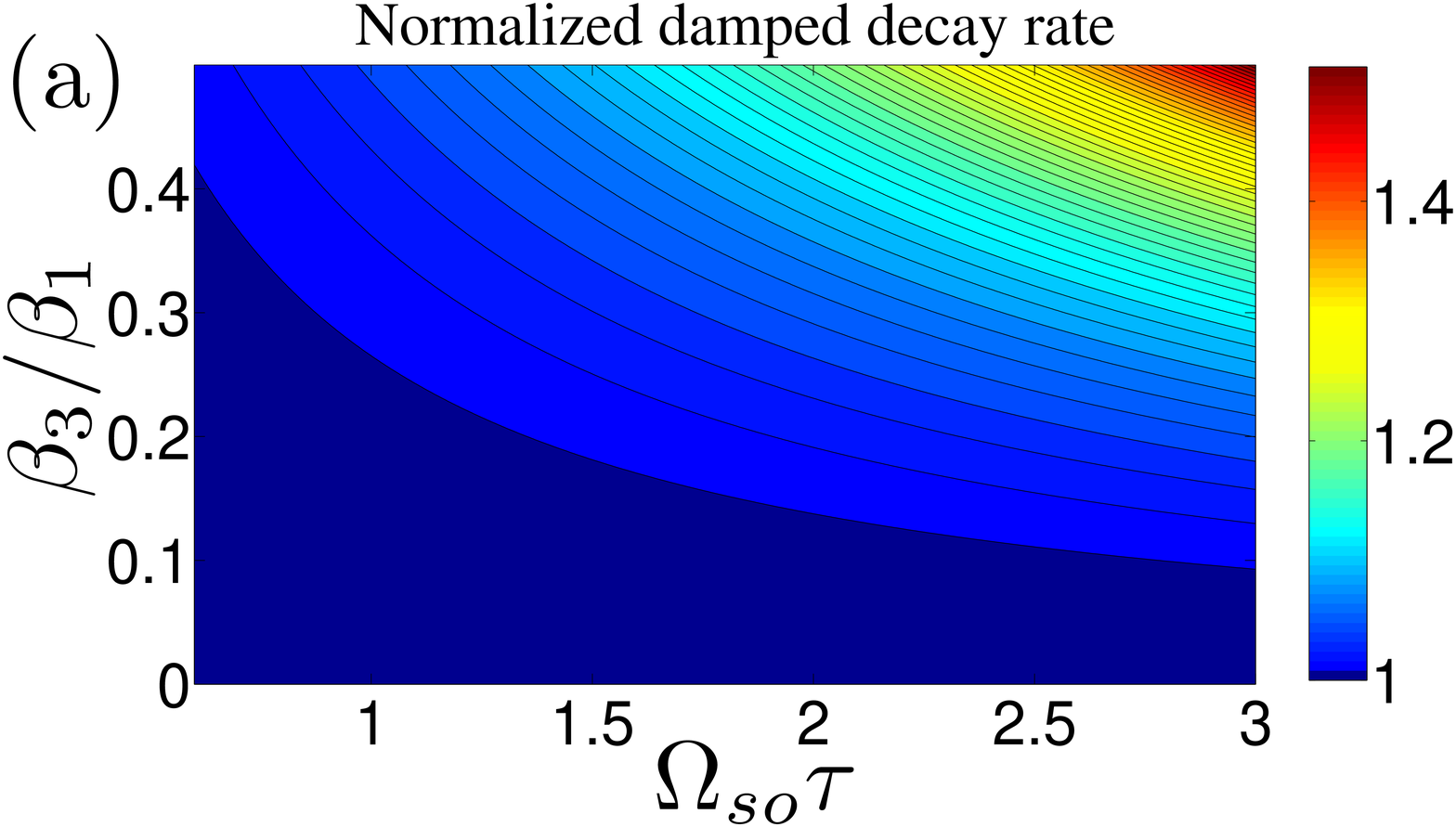}\\
\includegraphics[width=1.0\columnwidth]{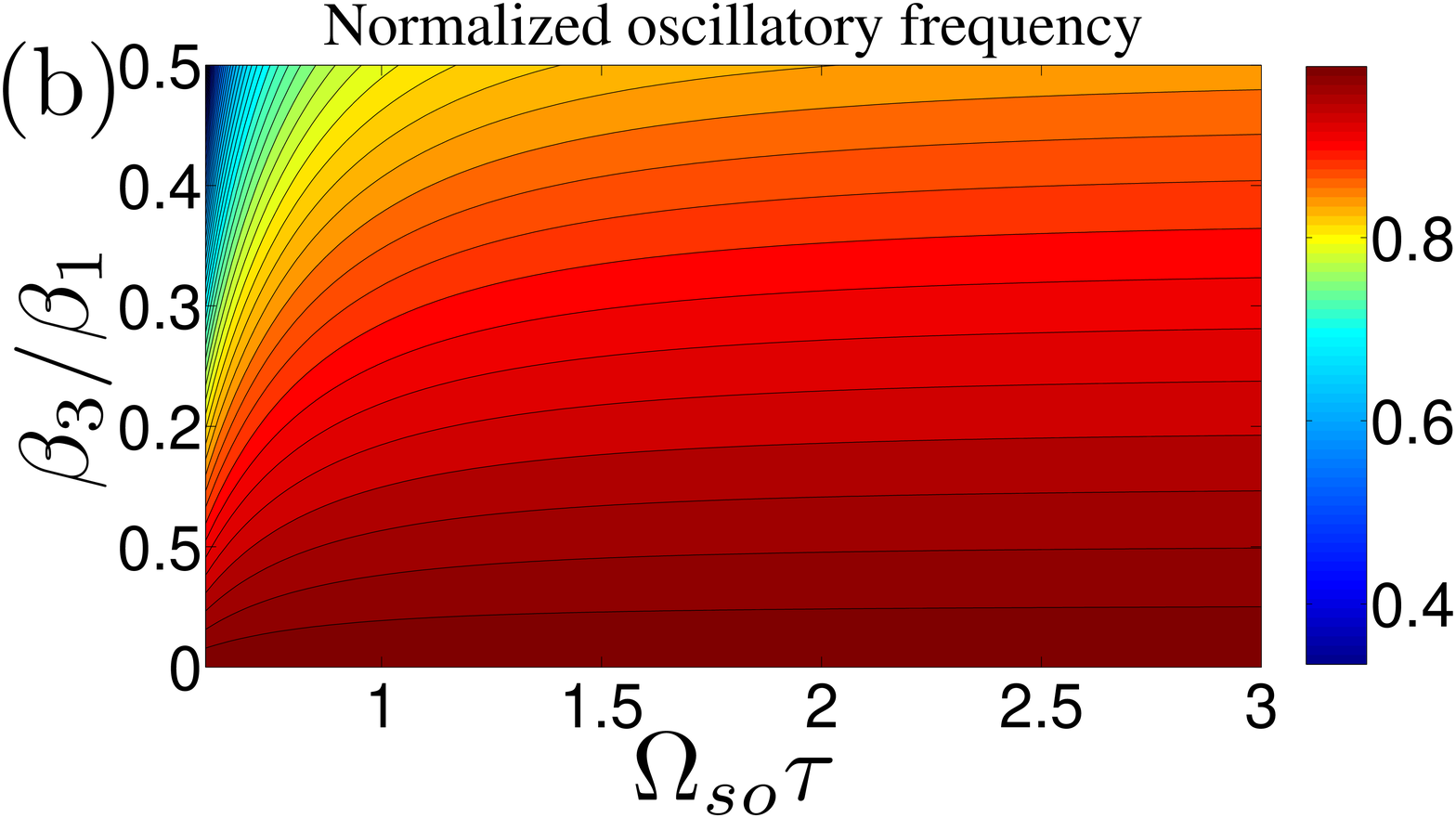}
\caption{The uniform spin dynamics from the weak to the strong spin-orbit coupling regime in the presence of linear $\beta_1$ and cubic $\beta_3$ SOI.  (a) The normalized exponential decay rate, $\Re(i\Omega\tau)$ is constant  when  $\beta_3$ is zero and slightly larger than $\frac{1}{2}$ when $\beta_3$ is nonzero. (b) The nonzero normalized oscillatory frequency, $\Im(i\Omega\tau)$, appear when $\Omega_{so}\tau > \frac{1}{2}$.}
\label{disp}
\end{figure}

Now, let us choose $\alpha=\beta_1$ which is a more unique case and gives us the persistent spin helix for special $q$ values.\cite{prl2007,B&H,psh} For the uniform spin polarization, the decay rate of the spin satisfies
\begin{eqnarray}\label{diffuson-2}
i\Omega\tau=1-\sqrt{1-2(\Omega_{so}\tau)^2},
\end{eqnarray}
where $\Omega_{so}=2\sqrt{\alpha^2+\beta_1^2}k_f$. The damped oscillation mode will happen when
$\Omega_{so}\tau=2\sqrt{2}\alpha k_f\tau>\sqrt{2}/2$, say $2\alpha k_f\tau>1/2$ which is the same as
 the pure Rashba or Dresselhauss case. The oscillating frequency in the clean limit is
$\sqrt{2}\Omega_{so}=4\alpha k_f$ which is the two fold of the frequency for the pure
Rashba or Dresselhauss case. On the other hand, as the real part of $i\Omega\tau$ is equal to $1$
when damped oscillation mode appear, the damped decay rate is also the two fold of the case of the pure Rashba or Dresselhauss.

\section{Spin dynamics and Rabi oscillation}
Before we discuss the spin dynamics for the nonuniform spin polarization system, let us give a physical explanation of the result we have obtained. We can construct a simple physics picture to describe the spin polarized wave theoretically. Taking the Rashba SOI for example, we define the eigenstates $|\phi_{k}^a\rangle$ to denote the majority band and the $|\phi_{k}^b\rangle$ to denote the minority band. The spin of the eigenstate of the SOC 2DEG lies in the $x-y$ plane. The majority electron has opposite spin to the minority electron when they have the same wave vector $k$.
\begin{figure}[htbp]
\includegraphics[width=0.8\columnwidth]{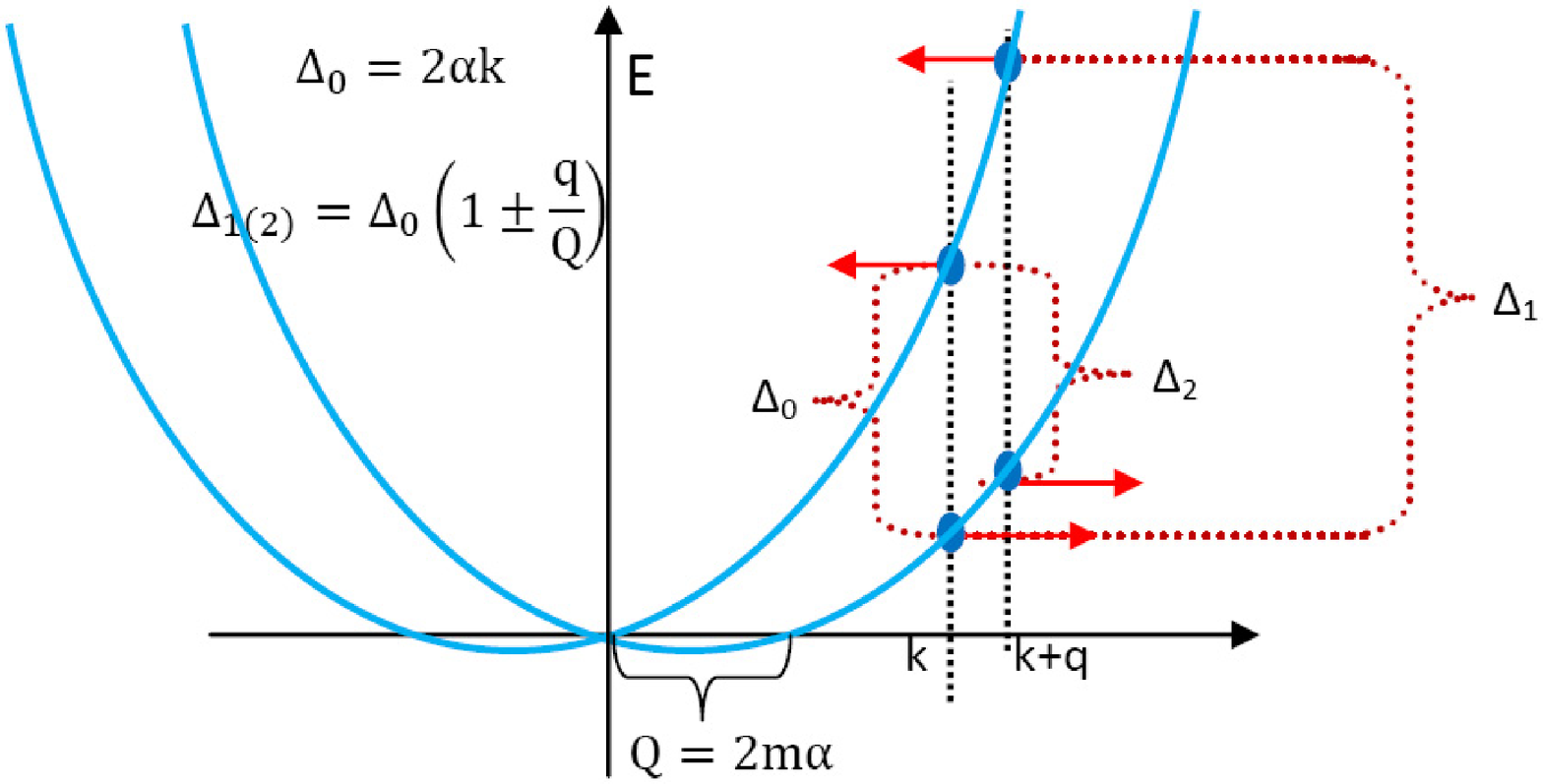}
\caption{The dispersion relation due to the linear Dresselhaus SOI. The SOI induces the energy gap $\Delta_0=2\beta_1 k$ which is the spin precession frequency for the single electron spin. However, when the system is excited to be a spin polarization wave with wave vector $q$, the spin polarization along the $z$ direction is constructed by the superposition of the two electron with wave vectors $k$ and $k+q$. In this case, the spin precession frequency will be $\Delta_{1(2)} \simeq \Delta_0(1\pm\frac{q}{Q})$, where $Q=2m\beta_1$.}
\label{spincouple}
\end{figure}
As a result, The spin polarization along the $z$ direction can be obtained by the superposition of the majority and minority bands as
\begin{eqnarray}\label{spinwave-1}
\psi_{\uparrow,q}&=&A[\sum_{k}e^{(\epsilon-\epsilon_f)^2/4\sigma^2}\frac{1}{\sqrt{2}}(|\phi_{k}^a\rangle+|\phi_{k+q}^b\rangle)]\nonumber \\&+&A[\sum_{k}e^{(\epsilon-\epsilon_f)^2/4\sigma^2}\frac{1}{\sqrt{2}}(|\phi_{k}^b\rangle+|\phi_{k+q}^a\rangle)],
\end{eqnarray}
where $A$ is the normalization coefficient, $\psi_{\uparrow,q}$ is the wave function of the system with positive spin polarization along $z$ direction with wave vector $q$ and the function $e^{(\epsilon-\epsilon_f)^2/4\sigma^2}$ restrict the spin polarization electrons only in the narrow range $\frac{1}{2\sigma} \ll \epsilon_f$ around the Fermi energy $\epsilon_f$. The expectation value $\langle\psi_{\uparrow,q}|\sigma_z\cos{q'x}|\psi_{\uparrow,q}\rangle$ is nonzero only when $q'=q$ which confirms that $\psi_{\uparrow,q}$ can describe the spin polarized wave. The energy difference of these two electrons in the first(second) term on the right hand side of Eq (\ref{spinwave-1})is $\Delta_{1(2)}$ as shown in Fig. (\ref{spincouple}). Therefore, $|\psi\rangle$ can be treated as a collective two level system with two Rabi frequencies $\Omega_{1(2)}=\frac{\Delta_{1(2)}}{\hbar}$. For the uniform spin polarization means $q=0$ and there is only one Rabi frequency $\Omega_0=\frac{\Delta_0}{\hbar}$ Fig.~\ref{spincouple}. When the system is very clean, our results, Eq.(\ref{puresoi},\ref{diffuson-2}), show that the spin evolution is damped oscillation and the oscillatory frequency is the Rabi frequency. It is a little surprising that when $\alpha=\beta$, although the SOC gap $\Delta_0$ is not a constant, the oscillatory frequency is corresponding to the maximum splitting energy $4\alpha k_f$ instead of the average splitting energy $2\sqrt{2}\alpha k_f$. In the weak SOC regime, the disorder is so strong that the splitting energy due to the SOI is completely submerged in the broadening of the band $\frac{\hbar}{\tau}$. Therefore, the spin polarization just exponential decays. For the non-uniform spin polarization case, since there are two Rabi oscillation frequencies $\Omega_{1}$ and $\Omega_{2}$, we expect to have two damped oscillatory modes in the clean system corresponding to energy differences $\Delta_{1}$ and $\Delta_2$ respectively in Fig.~\ref{spincouple}.

\section{Non-uniform spin polarization}
\begin{figure}[ht]
\includegraphics[width=0.8\columnwidth]{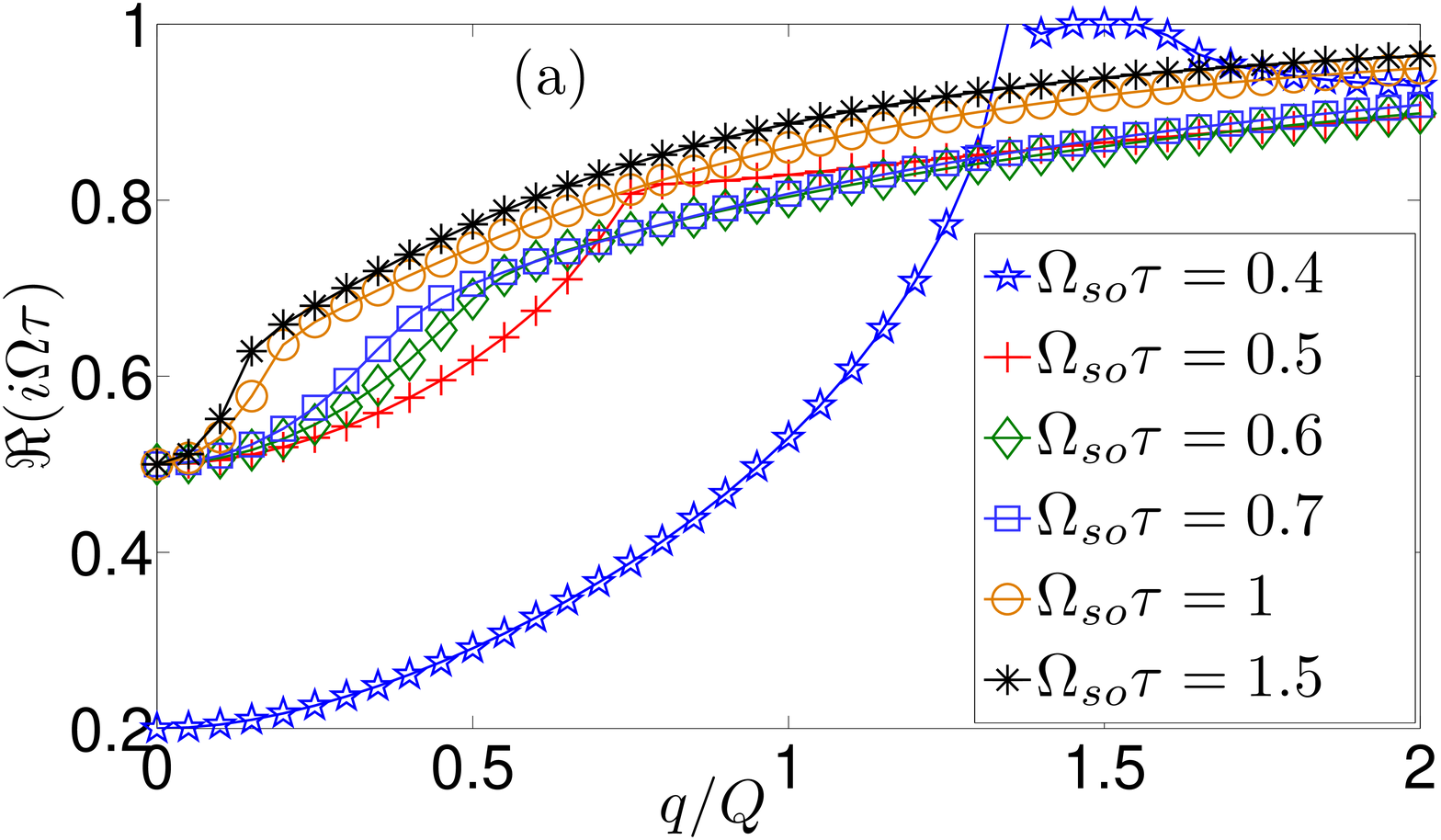}\\
\includegraphics[width=0.8\columnwidth]{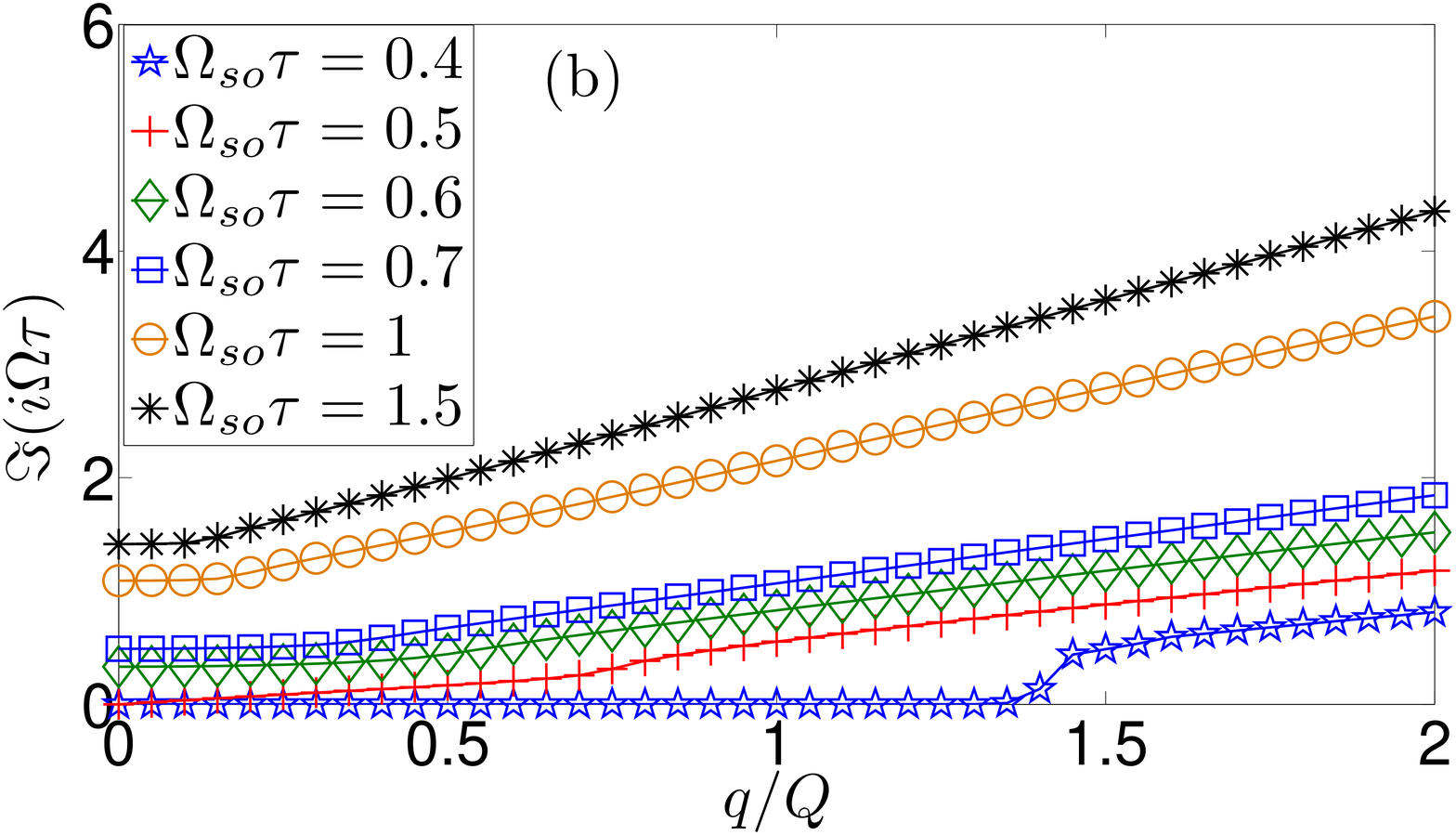}
\caption{The fast oscillatory mode of the nonuniform spin dynamics in the strong SOC regime when the system only has bulk inversion asymmetry. (a) The normalized exponential decay rate, $\Re(i\Omega\tau)$ increase with increasing $q$ and approach to one at large $q$. (b) The nonzero normalized oscillatory frequency, $\Im(i\Omega\tau)$, increases linearly at large $q$, the slope is close to $\Omega_{so}\tau$ and its value approaches $\Omega_{so}(1+\frac{q}{Q})$.}
\label{faster}
\end{figure}

\begin{figure}[ht]
\includegraphics[width=0.8\columnwidth]{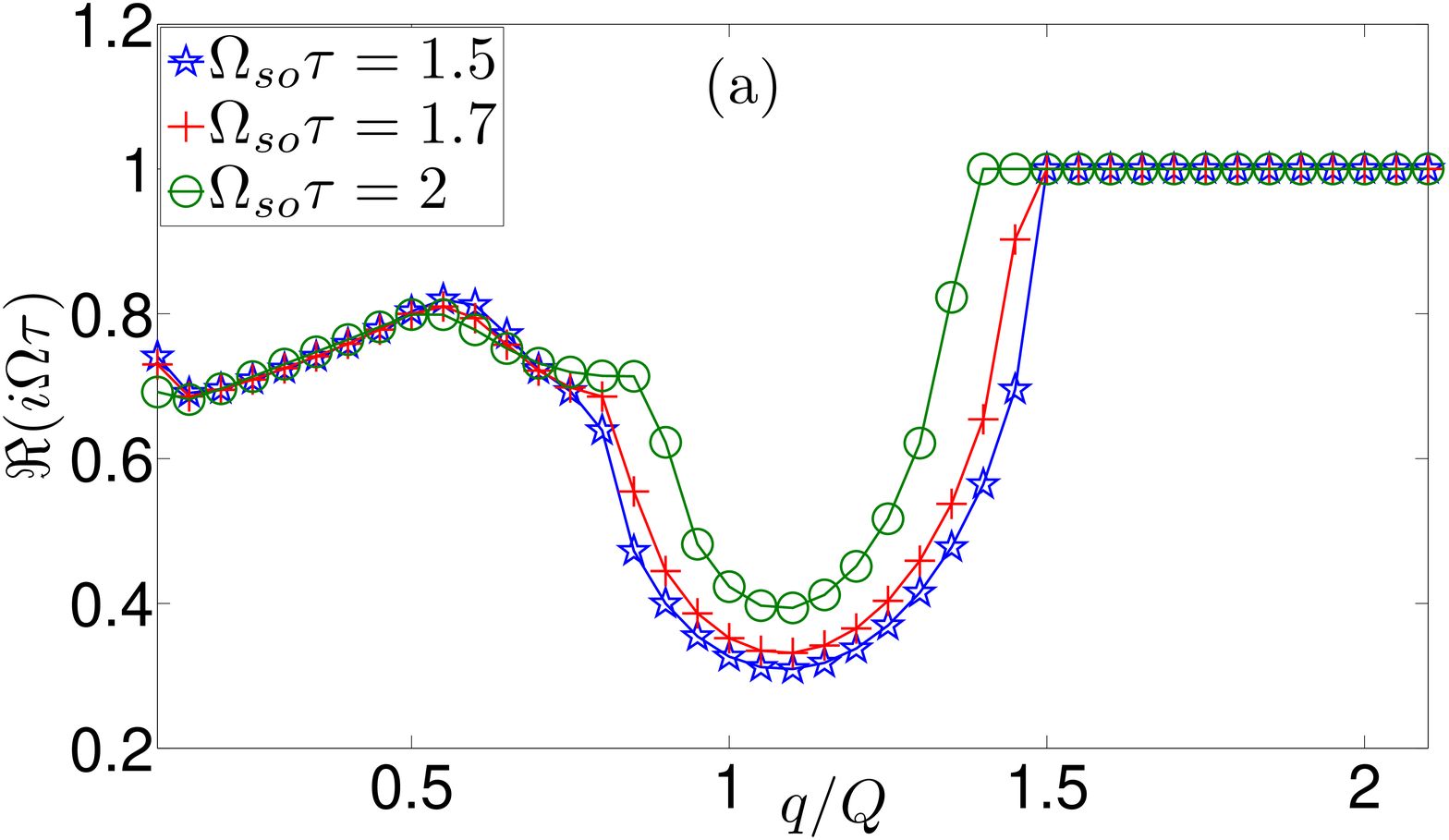}\\
\includegraphics[width=0.8\columnwidth]{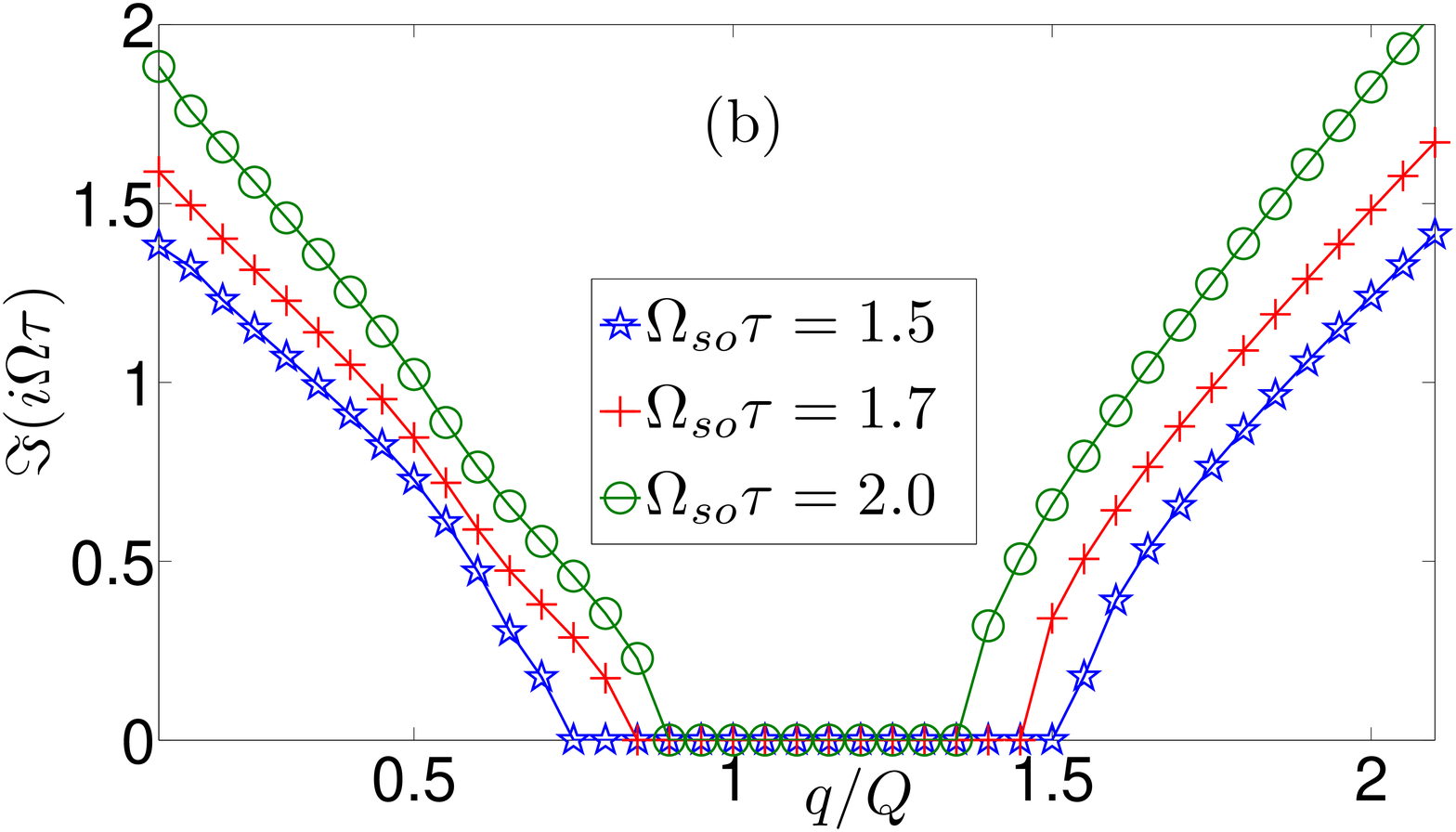}
\caption{The slow oscillatory mode of the nonuniform spin dynamics in the strong SOC regime when the system only has bulk inversion asymmetry. (a) The normalized exponential decay rate, $\Re(i\Omega\tau)$ has a minimum around $q=Q$ and approach to one at large $q$. (b) The nonzero normalized oscillatory frequency, $\Im(i\Omega\tau)$ is always zero when $q$ is around $Q$ and increases linearly at large $q$. The slope is close to $\Omega_{so}\tau$ and the value approaches $\Omega_{so}(1-\frac{q}{Q})$ at large $q$.}
\label{slower}
\end{figure}

In the case of the non-uniform spin polarized 2DEG, the initial state is a spin wave with wave vector $q$, the momentum $\mathbf{k}$ is coupled to $\mathbf{k}+\mathbf{q}$ which makes the center of the Fermi sea be shifted to near $\mathbf{q}$.
 the average magnetic field is nonzero and the off diagonal elements of the diffusion matrix appear to couple the different spin component.  When only considering the Rashba or linear Dresselhaus SOI, our numerical calculation does have two kinds of spin dynamical modes which are shown in Fig.(\ref{faster},\ref{slower}).

The damped oscillatory modes at large $q$ can be approximately written as
\begin{eqnarray}
i\Omega\tau=1-\sqrt{1-(\Omega_{so}\tau)^2(1\pm\frac{q}{Q})^2}
\label{decay-1}
\end{eqnarray}
where $Q=2m\lambda_1$.

The two damped oscillatory mode and their oscillatory frequency satisfy our expectation based on the Rabi oscillation viewpoint. When $q$ increases, the Rabi frequency of the faster mode always increases which makes the damped oscillatory mode  appear even when $\Omega_{so}\tau<\frac{1}{2}$. This means we can expect to observe the oscillation for the nonzero spin polarization at higher temperature than for the uniform spin polarization. In Ref. \onlinecite{exq=0} where the spin polarization is uniform, the damped oscillatory mode appears below $5$ K. On the other hand in Ref. \onlinecite{nature2005}, where the spin polarization is nonuniform, the damped oscillatory mode appears below $50$ K. The material, Fermi energy and mobility in these two papers are similar. This seems support our Rabi oscillation viewpoint.  For the slow oscillatory mode, when $q$ is around $Q$, the corresponding Rabi frequency $\Omega_{2}$ is around $0$ which means the spin precession is very slow. Because the Rabi frequencies is much smaller than $\frac{1}{\tau}$, the spin polarization just decays exponentially and the exponential decay rate has its minimum in this regime when $q$ is around $Q$ . A particular case is when $\alpha=\beta_1$, $q=\pm 4m\alpha$ and along the $x$ direction and $\beta_3=0$ defined in Eq.(\ref{ha}), the Rabi frequency of the slower mode is zero for all of the electron momentum $k$ and  Eq.(\ref{decay-1}) becomes the exact solution.\cite{B&H} On the other hand, the spin $y$ is a good quantum number for all the electron states which means the spin independent disorder will never couple the two electrons in different bands with different spin directions. Therefore, the Rabi frequency of the slower mode is still exactly zero even in the presence of the spin independent disorder no matter how strong it is. As a result, the spin along the $z$-direction will never precess and has infinite long life time. This provides another way to understand the persistent spin helix.\cite{prl2007,psh} However, the cubic Dresselhaus SOI induces band transition in the presence of spin independent impurities and makes the spin life time finite.\cite{a+b}

\section{Proposed experiments}
The spin dynamics in the strong SOC regime have several special characters which can be used in experimental measurements.

\textit{Momentum scattering time $\tau_p^*$}:    In the spin dynamics, the Coulomb interaction plays an important role in determining the momentum scattering time $\tau_p^{*}$ \cite{e-e-1,e-e-2}. This is quite different to the charge transport case where electron-electron(e-e) interaction will not change the ensemble momentum scattering $\tau_p$ which determines the electron mobility. This difference is called spin Coulomb drag (SCD). In  previous experimental work, SCD was observed through  the spin diffusion coefficient $D_s=\frac{1}{2}v_f^2 \tau_p^*$ by fitting the spin decay rate in the weak SOC regime. Here, we provides a way to observe SCD in the strong SOC regime by directly measuring the momentum scattering time $\tau_p^*$.  Based on Eq \ref{puresoi},\ref{izz-cubic-6}, when only Dresselhaus SOI is presented, the damped decay rate is always almost equal to  $1/2$ as long as $\frac{\beta_3}{\beta_1}<0.2$ which is easily  realized in experiments.\cite{exq=0-1,psh}

\textit{The strength of SOIs}:  Here, we would like to emphasize that $2\beta_1 k_f \tau=\frac{1}{2}$ is a very important case and is corresponding to the transition point between pure exponential decay mode and damped oscillatory mode.  The decay rate at this point is not only equal to $\frac{1}{2\tau}$, but also equal to $\frac{1}{2\beta_1 k_f}$ when $\alpha=0$. This means that at this point we can obtain the strength of linear Dresselhaus SOI from the spin polarization decay rate.  When $2\beta_1 k_f\tau=\frac{1}{2}$, we can increase the Rashba SOI by adding a gate voltage. As long as $0<\alpha<\beta_1$, according to Eq. \ref{zmode}, the spin evolution is still  decay exponentially and the decay rate is $(1-\frac{\sqrt{2}}{2}\sqrt{1-2(\Omega_{so}\tau)^2})/\tau$ where $\Omega_{so}=2\sqrt{\alpha^2+\beta_1^2} k_f$ which gives us the strength of Rashba SOI. 

\section{Conclusion}
We have discussed the spin dynamics in the strong spin-orbit coupling regime. We describe quantitatively the special characters of the damped oscillatory mode in this regime. We also compare our result to the previous experimental data and find they match very well. Based on our theoretical results, a reliable way is proposed to measure the Rashba and Dresselhaus coefficients and electron momentum scattering time which is not corresponding to the mobility due to the Coulomb interaction. Furthermore, we find that the spin dynamics in the 2DEG can be treated as a collective two level system. This helps us semi-quantitatively understand the spin dynamics in the strong spin-orbit coupling regime. For the nonzero spin polarization case, we predict that there exist double damped oscillatory modes at large $q$ and explain the persistent spin helix mode from the Rabi oscillation point of view.

We acknowledge Chia-Ren Hu, Ar. Abanov, Yang Liu and Victor Galitski for very helpful discussion and support from DMR-0547875, NSF-MRSEC DMR-0820414, and SWAN-NRI. J. S. is a Cottrell
Scholar of Research Corporation.

\appendix
\begin{widetext}
\section{Spin dynamic matrix for the uniform spin polarization}
In this section, we derive the  spin evolution mode of the uniform spin polarization. According to Eq. \ref{hso} the strength of SOI is angle dependent and can be written as
\begin{eqnarray}\label{hso-2}
h_{so}=\sqrt{\alpha^2+\beta_1^2} k_f \sqrt{1+\cos2\psi\cos2\theta+(2(\frac{\beta_3}{\lambda'})^2-\frac{2\beta_3}{\lambda'}\sin(\psi+\pi/4))(1+\cos4\theta)-\frac{4\beta_3}{\lambda'}\cos(\psi+\pi/4)\cos2\theta}
\end{eqnarray}
where $\cos\psi=\lambda_1/\sqrt{\lambda_1^2+\lambda_2^2}$.

First we consider the case for $\beta_3=0$. The Hamiltonian is written as
\begin{eqnarray}\label{hamab}
H&=&\frac{k^2}{2m}+(\alpha+\beta)k_x\sigma_y-(\alpha-\beta)k_y\sigma_x,\nonumber \\
&=&\frac{k^2}{2m}+\lambda_1 k_x\sigma_y+\lambda_2 k_y\sigma_x,
\end{eqnarray}
where $k_x$ is along the $[1 1 0]$ direction, $\lambda_1=\alpha+\beta$ and $\lambda_2=-(\alpha-\beta)$. The Green's function for this Hamiltonian takes the form
\begin{eqnarray}\label{greena+b}
G^{R(A)}&=&\frac{E-\frac{k^2}{2m}\pm \frac{i}{2\tau}+(\alpha+\beta)k_x\sigma_y-(\alpha-\beta)k_y\sigma_x}{(E-\frac{k^2}{2m}\pm \frac{i}{2\tau})^2-(\alpha^2+\beta^2)k^2(1+\frac{2\alpha\beta}{\alpha^2+\beta^2}\cos2\theta)}\nonumber \\
&=&\frac{E-\frac{k^2}{2m}\pm \frac{i}{2\tau}+\lambda_1 k_x\sigma_y+\lambda_2 k_y\sigma_x}{(E-\frac{k^2}{2m}\pm \frac{i}{2\tau})^2-\frac{(\lambda_1^2+\lambda_2^2)}{2}k^2(1+\gamma\cos2\theta)},
\end{eqnarray}
where $\tau$ is the momentum scattering time, $\gamma=\frac{2\alpha\beta}{\alpha^2+\beta^2}=\frac{\lambda_1^2-\lambda_2^2}{\lambda_1^2+\lambda_2^2}$. It is more convenient to write down the element of the $2\times 2$ Green's function Eq.(\ref{greena+b}) as
\begin{eqnarray}\label{grennele}
G_{11}&=&G_{22}=\frac{1}{2}(\frac{1}{E-\frac{k^2}{2m}-\lambda k\sqrt{1+\gamma\cos2\theta}\pm\frac{i}{\tau}}+\frac{1}{E-\frac{k^2}{2m}+\lambda k\sqrt{1+\gamma\cos2\theta}\pm\frac{i}{\tau}}),\nonumber\\
G_{12}&=&\frac{1}{2}(\frac{1}{E-\frac{k^2}{2m}-\lambda k\sqrt{1+\gamma\cos2\theta}\pm\frac{i}{\tau}}-\frac{1}{E-\frac{k^2}{2m}+\lambda k\sqrt{1+\gamma\cos2\theta}\pm \frac{i}{\tau}})\frac{\sqrt{2}(-i\cos\psi\cos\theta+\sin\psi\sin\theta)}{\sqrt{1+\gamma\cos2\theta}}\nonumber \\
G_{21}&=&\frac{1}{2}(\frac{1}{E-\frac{k^2}{2m}-\lambda k\sqrt{1+\gamma\cos2\theta}\pm \frac{i}{\tau}}-\frac{1}{E-\frac{k^2}{2m}+\lambda k\sqrt{1+\gamma\cos2\theta}\pm \frac{i}{\tau}})\frac{\sqrt{2}(i\cos\psi\cos\theta+\sin\psi\sin\theta)}{\sqrt{1+\gamma\cos2\theta}}
\end{eqnarray}
where $\lambda=\sqrt{(\lambda_1^2+\lambda_2^2)/2}=\sqrt{\alpha^2+\beta^2}$, $\cos\psi=\lambda_1/\sqrt{\lambda_1^2+\lambda_2^2}$ and $\gamma=\cos2\psi$.

According to Eq.(\ref{Ics}), the diagonal element of the spin polarization along $z$ direction has the form
\begin{eqnarray}\label{Izz-app}
&&I^{zz}=I_{11,11}-I_{11,22}-I_{22,11}+I_{22,22}\nonumber \\
&&=\frac{1}{2m\tau}\int \frac{d^2k}{(2\pi)^2} (G_{11}^AG^R_{11}-G^A_{21}G^R_{12}-G^A_{12}G^R_{21}+G_{22}^{A}G_{22}^R).
\end{eqnarray}

The first term and the fourth term in Eq. \ref{Izz-app} are equal  to each other and have the form 
\begin{eqnarray}
&&\frac{1}{2m\tau}\int\frac{d^2k}{(2\pi)^2}G^A_{11}G^R_{11}=\frac{1}{2m}\int\frac{d^2k}{(2\pi)^2}\frac{1}{4}(\frac{1}{E-\epsilon_+(k)-\frac{i}{2\tau}}+\frac{1}{E-\epsilon_-(k)-\frac{i}{2\tau}})\times \nonumber \\
&&\ \ \ \ \ \ \ \ \ \ \ \ \ \ \ \ \ \ \ \ \ (\frac{1}{E+\Omega-\epsilon_-(k)+\frac{i}{2\tau}}+\frac{1}{E+\Omega-\epsilon_-(k)+\frac{i}{2\tau}}).\nonumber
\end{eqnarray}
\begin{eqnarray}\label{I1111}
&&=\frac{1}{16m\pi}\int_{0}^{2\pi}\frac{d\theta}{v_f}(\frac{k^+}{1-i\Omega\tau}+\frac{k^-}{1-i\Omega\tau+2i\lambda k\sqrt{1+\gamma\cos2\theta}}+\frac{k^+}{1-i\Omega\tau-2i\lambda k\sqrt{1+\gamma\cos2\theta}}+\frac{k^-}{1-i\Omega\tau}),\nonumber \\
&&\simeq \frac{1}{16\pi}\int_{0}^{2\pi} d\theta(\frac{1}{1-i\Omega\tau}+\frac{1}{1-i\Omega\tau+2i\lambda k\sqrt{1+\gamma\cos2\theta}}+\frac{1}{1-i\Omega\tau-2i\lambda k\sqrt{1+\gamma\cos2\theta}}+\frac{1}{1-i\Omega\tau}),\nonumber \\
\end{eqnarray}
where $v_f=\frac{\partial E_f}{\partial k}$.
In the polar coordinate, $\int d^2k=\int d(k^2/2)d\theta$. As we assume that $\lambda k_f\ll E_f$, $d(k^2/2)\simeq mdE$ where $m$ is the effective mass.

The other two terms are also equal to each other and can be written as
\begin{eqnarray}\label{I1212}
&&\frac{1}{2m\tau}\int\frac{d^2k}{(2\pi)^2}G^A_{21}G^R_{12}=\int\frac{d^2k}{(2\pi)^2}\frac{1}{4}(\frac{1}{E-\epsilon_+(k)-\frac{i}{2\tau_e}}-\frac{1}{E-\epsilon_-(k)-\frac{i}{2\tau_e}})\times \nonumber \\
&&\ \ \ \ \ \ \ \ \ \ \ \ \ \ \ \ \ \ \ \ \ (\frac{1}{E+\Omega-\epsilon_+(k)+\frac{i}{2\tau_e}}-\frac{1}{E+\Omega-\epsilon_-(k)+\frac{i}{2\tau_e}})\nonumber \\
&&=\frac{1}{16m\pi}\int_{0}^{2\pi}\frac{d\theta}{v_f}(\frac{k^+}{1-i\Omega\tau}-\frac{k^-}{1-i\Omega\tau+2i\lambda k\sqrt{1+\gamma\cos2\theta}}-\frac{k^+}{1-i\Omega\tau-2i\lambda k\sqrt{1+\gamma\cos2\theta}}+\frac{k^-}{1-i\Omega\tau}).\nonumber \\
&&\simeq \frac{1}{16\pi}\int_{0}^{2\pi} d\theta(\frac{1}{1-i\Omega\tau}-\frac{1}{1-i\Omega\tau+2i\lambda k\sqrt{1+\gamma\cos2\theta}}-\frac{1}{1-i\Omega\tau-2i\lambda k\sqrt{1+\gamma\cos2\theta}}+\frac{1}{1-i\Omega\tau},\nonumber \\
\end{eqnarray}
Substituting Eq. \ref{I1111} ,\ref{I1212} to Eq. \ref{Izz-app}, we have
\begin{eqnarray}\label{Izz-app}
&&I^{zz}=\frac{1}{4\pi}\int_{0}^{2\pi} d\theta (\frac{1}{1-i\Omega\tau+2i\lambda k\sqrt{1+\gamma\cos2\theta}}+\frac{1}{1-i\Omega\tau-2i\lambda k\sqrt{1+\gamma\cos2\theta}})\nonumber \\
&&=\frac{1}{2\pi}\int_{0}^{2\pi} d\theta \frac{1-i\Omega\tau}{(1-i\Omega\tau)^2+(\Omega_{so}\tau)^2(1+\gamma\cos2\theta)}\nonumber \\
&&=\frac{1-i\Omega\tau}{2\pi((1-i\Omega\tau)^2+(\Omega_{so}\tau)^2)}\int_{0}^{\pi} dx \frac{2}{(1+a\cos(x))},
\end{eqnarray}
where $x=2\theta$ and $a=\gamma(\Omega_{so}\tau)^2/((1-i\Omega\tau)^2+(\Omega_{so}\tau)^2)$. 
 The indefinite integral $\int dx\frac{1}{1+a\cos(x)}=\frac{2 arc\tanh \left[\frac{(-1+a ) \tan\left[\frac{x}{2}\right]}{\sqrt{-1+a^2}}\right]}{\sqrt{-1+a^2}}$. Therefore we have
\begin{eqnarray}\label{spdiff}
I^{zz}&=&\frac{1-i\Omega\tau}{2\pi((1-i\Omega\tau)^2+(\Omega_{so}\tau)^2)}\int_{0}^{\pi} dx \frac{2}{(1+a\cos(x))}\nonumber \\
&=&\frac{1-i\Omega\tau}{2\pi((1-i\Omega\tau)^2+(\Omega_{so}\tau)^2)}\times 2(\frac{2 arc\tanh \left[\frac{(-1+a ) \tan\left[\frac{\pi}{2}\right]}{\sqrt{-1+a^2}}\right]}{\sqrt{-1+a^2}}
-\frac{2 arc\tanh \left[\frac{(-1+a ) \tan\left[\frac{0}{2}\right]}{\sqrt{-1+a^2}}\right]}{\sqrt{-1+a^2}})\nonumber \\
&=&\frac{1-i\Omega\tau}{2\pi((1-i\Omega\tau)^2+(\Omega_{so}\tau)^2)}\times \frac{2\pi i}{\sqrt{-1+a^2}}=\frac{1-i\Omega\tau}{\sqrt{((1-i\Omega\tau)^2+(\Omega_{so}\tau)^2)^2-\gamma^2(\Omega_{so}\tau)^4}}.\nonumber \\
\end{eqnarray}

When Rashba SOI is zero, the strength of SOIs takes the form
\begin{eqnarray}\label{hso-3}
h_{so}=\beta_1 k_f\sqrt{1+(2\frac{\beta_3^2}{\beta_1^2}-2\frac{\beta_3}{\beta_1})(1+\cos4\theta)}.
 \end{eqnarray} 
To obtain the spin diffusive matrix element $I^{zz}$, it is easy to prove that we only need to replace the term $\lambda k\sqrt{1+\gamma\cos2\theta}$ in Eq. \ref{spdiff} with $h_{so}$ in Eq. \ref{hso-3}. Therefore, we have
 \begin{eqnarray}\label{izz-cubic-4}
I^{zz}&=&\frac{1}{4\pi}\int (\frac{2(1-i\Omega\tau)}{(1-i\Omega\tau)^2+(2 h_{so}\tau)^2})d\theta \nonumber \\
&=&\frac{1-i\Omega\tau}{\sqrt{(1-i\Omega\tau)^2+\Omega_{so}^2\tau^2}\sqrt{(1-i\Omega\tau)^2+\Omega_{so}^2\tau^2(1+2(\frac{\beta_3}{\beta_1})^2-2\frac{\beta_3}{\beta_1})}}.
\end{eqnarray}
\end{widetext}

\noindent


\end{document}